\documentclass[journal]{IEEEtran}
\usepackage{amsmath,amsfonts}
\usepackage{array}
\usepackage[caption=false,font=normalsize,labelfont=sf,textfont=sf]{subfig}
\usepackage{textcomp}
\usepackage{stfloats}
\usepackage{url}
\usepackage{verbatim}
\usepackage{graphicx}
\usepackage{xcolor}
\usepackage{hyperref}
\hypersetup{
    colorlinks=true,
    linkcolor=black,
    citecolor=black,
    urlcolor=black,
    breaklinks=true
}
\usepackage{algorithm}
\usepackage{float}
\usepackage{algpseudocode}
\usepackage{stfloats}
\usepackage{lipsum}
\usepackage{multirow}    
\usepackage{tabularx}    
\usepackage{makecell}    
\usepackage{placeins}
\usepackage{booktabs}
\usepackage{balance}
\usepackage{ifthen}
\usepackage{tikz}
\usepackage{adjustbox}
\usepackage{pgfplots}
\pgfplotsset{compat=1.18}
\usepackage{pifont}
\usepackage{orcidlink}
\newcommand{\orcidraise}[1]{\raisebox{0.5ex}{\scalebox{1.1}{\orcidlink{#1}}}}
\newcommand{\cmark}{\ding{51}} 
\newcommand{\xmark}{\ding{55}} 
\usepackage{Settings}
 
\begin{document}

\title{\fontsize{24pt}{28pt}\selectfont An Emerging NVM-Based On-Chip Training Architecture with Non-Ideality Mitigation Through Bipolar Weight Distributions}

\author{
    Peng Dang\orcidraise{0009-0006-5773-199X}, 
    Youna Huang\orcidraise{0009-0008-9409-279X}, 
    Yintao He\orcidraise{0000-0003-3054-0617}, 
    and Huawei Li\orcidraise{0000-0001-8082-4218}, \IEEEmembership{Senior Member,~IEEE}

\thanks{
Peng Dang, Yintao He and Huawei Li are with the State Key Laboratory of Processors, Institute of Computing Technology, Chinese Academy of Sciences, Beijing 100190, China, and also with the University of Chinese Academy of Sciences, Beijing 100190, China (e-mail: dangpeng21@mails.ucas.ac.cn; heyintao19z@ict.ac.cn; lihuawei@ict.ac.cn).

Youna Huang is with the Southern University of Science and Technology, Shenzhen 518066, China (e-mail: 12131007@mail.sustech.edu.cn). 
}}

\maketitle
\begin{abstract}
The rapid advancement of deep learning has presented significant energy efficiency challenges to the conventional von Neumann architecture. In-memory computing (IMC) architectures based on emerging non-volatile memory (eNVM) are widely regarded as a promising solution for accelerating neural network training due to their high parallelism and low power consumption.
However, the intrinsic non-idealities of eNVM devices can cause conductance updates to deviate from target values, thereby limiting the performance of on-chip training.
To address this challenge, this paper presents a Non-ideality Optimized eNVM Accelerator (NOVA) architecture for on-chip training. Specifically, we first fabricate a two-dimensional (2D) ferroelectric field-effect transistor (FeFET) and develop a conductance modulation behavioral model calibrated with experimental data. Building upon this device model, we propose, for the first time, a Non-ideality Avoidance Training (NAT) algorithm tailored for eNVM devices, which mitigates accuracy degradation by guiding weight convergence toward the most stable conductance regions of eNVM devices. 
Experimental results demonstrate that, even under severe device asymmetry, NAT improves the accuracy by an average of 15.1\% over the baseline methods across multiple benchmark tasks. Meanwhile, the NOVA achieves an average energy efficiency gain of approximately 33.58$\times$ compared with the peak energy efficiency of graphics processing units (GPUs).
\end{abstract}
\section{Introduction}
\IEEEPARstart{T}{he} explosive growth of artificial intelligence (AI) has led to the emergence of increasingly complex neural network models, imposing unprecedented demands on computational capability and data throughput. However, conventional von Neumann architectures, which are characterized by the physical separation between compute units and memory, are fundamentally constrained by the ``memory wall'' bottleneck. This limitation arises from the frequent movement of data between memory and compute units, resulting in significant latency and energy overhead~\cite{mehonic2022brain}. To overcome this challenge, IMC architectures based on emerging non-volatile memory (eNVM) technologies such as resistive random-access memory (ReRAM), ferroelectric field-effect transistors (FeFETs), and magnetic random-access memory (MRAM) have emerged as a promising paradigm~\cite{zhang2020neuro}. By embedding computation directly within the memory array, IMC architectures drastically reduce data movement and leverage the inherent parallelism of memory arrays to efficiently execute large-scale matrix-vector multiplication (MVM) operations~\cite{10375354}.

Currently, most eNVM-based accelerators support only inference~\cite{9731725,9830153,10067544,yao2020fully,3676792}. With continuous improvements in programming precision and endurance of eNVM devices, on-chip training architectures that enable end-to-end training are becoming increasingly feasible~\cite{9062979,zhang2023edge,cai2019fully,wan2022compute,ambrogio2018equivalent}. In on-chip training, forward propagation, backpropagation, and weight update of neural networks are performed within the eNVM array, which significantly reduces energy consumption and latency overhead associated with external memory accesses and data movement. Moreover, on-chip training enables hardware-aware training under real hardware conditions, endowing the model with the ability to continuously adapt and learn in the presence of device non-idealities~\cite{Diware}.

Although on-chip training is theoretically feasible, existing eNVM-based accelerators still face significant challenges in efficiently supporting it:

\textbf{First Challenge: Nonlinear and asymmetric conductance tuning.}  
In eNVM-based on-chip training, ideal weight updates require a linear relationship between the conductance change and the desired weight adjustment. However, the conductance modulation of eNVM cells typically exhibits high nonlinearity (NL) and asymmetry~\cite{rasch2023hardware,wan2022compute}. These non-ideal characteristics cause identical weight updates to be mapped to a conductance state that deviates from the intended target, degrading training accuracy~\cite{9292971,9716051}.

\textbf{Second Challenge: Inherent stochasticity and variability.}  
In addition to nonlinearity, eNVM devices suffer from cycle-to-cycle (C2C) fluctuations and device-to-device (D2D) variations during write operations~\cite{10070583,9775004}. In large-scale eNVM arrays, this unpredictable stochasticity accumulates progressively, severely compromising the precision of weight updates and undermining both training stability and final model accuracy~\cite{rasch2023hardware,9292971}.

Most existing research on IMC-based on-chip training accelerators largely overlooks the critical non-ideal characteristics of eNVM devices~\cite{8333741,ankit2020panther,August2017,9139434,10753271}. Although these works report near-software-baseline accuracy, their evaluations are substantially decoupled from realistic hardware behavior, rendering the reported results difficult to reproduce in practice. In contrast, several recent studies~\cite{rasch2023hardware,9292971,9716051,lim2019adaptive,10233046,yi2023activity,102206,abm8537} attempt to incorporate more physically realistic non-ideal models to improve evaluation fidelity. Table~\ref{tab:Weight_Comparison} summarizes the key characteristics of these related works.
The studies in~\cite{102206,abm8537} develop highly linear eNVM devices through material and process optimization to mitigate hardware non-idealities at their source. While effective, this approach relies heavily on specific advanced fabrication processes and device properties. The studies in~\cite{rasch2023hardware,9292971,yi2023activity,9716051} perform noise-aware training by incorporating non-ideality models into the training process, leveraging the inherent robustness of neural networks to passively tolerate device non-idealities. However, the effectiveness of these methods is highly dependent on the linearity of the eNVM devices; as linearity degrades, convergence difficulties and significant accuracy degradation often arise. This situation underscores the necessity of an algorithm-device co-design paradigm that explicitly incorporates the physical characteristics of eNVM devices at the algorithmic level.

To address the aforementioned challenges, we propose NOVA, an on-chip training accelerator architecture compatible with multiple emerging memory technologies. NOVA employs our fabricated FeFET devices as the core computing unit to enable on-chip training. To mitigate the impact of device non-idealities on training performance, we further propose the NAT algorithm. As shown in Table~\ref{tab:Weight_Comparison}, unlike existing methods, NAT actively guides model weights toward conductance-stable regions of eNVM devices during training, thereby reducing weight-to-conductance mapping errors. Specifically, the main contributions of this work are summarized as follows:

\begin{itemize}{}{}
    \item We design an eNVM-based on-chip training architecture (NOVA) that efficiently supports end-to-end on-chip training of neural networks. This architecture enables flexible dataflow switching during on-chip training by reusing a single set of peripheral circuits, significantly improving training efficiency.

    \item We fabricate FeFET devices based on a CuInP\textsubscript{2}S\textsubscript{6}/MoS\textsubscript{2} heterostructure and perform curve fitting and modeling using experimental data. The resulting model accurately captures the dynamic conductance modulation behavior of the devices and effectively emulates their real-world response during on-chip training.

    \item We propose, for the first time, a training algorithm tailored to mitigate the impact of eNVM device non-idealities, termed NAT. By steering model weights away from conductance regions where device non-idealities are most pronounced, NAT substantially enhances both training stability and final model accuracy.
    
    \item 
    Experimental results demonstrate that, even under severe device asymmetry, NAT improves the training accuracy by an average of 15.1\% over baseline methods across multiple benchmark tasks. Meanwhile, NOVA achieves an average energy efficiency improvement of 33.58$\times$ compared with the peak energy efficiency of GPUs.
\end{itemize}

The remainder of this paper is organized as follows. Section \ref{sec:section_2} presents related work. Section \ref{sec:section_3} describes the implementation details of NOVA. Section \ref{sec:section_4} presents experimental results using VGG6, VGG11, and ResNet18, and analyzes the performance of on-chip training with the NOVA architecture. Section \ref{sec:section_5} concludes the paper.

\newcolumntype{C}[1]{>{\centering\arraybackslash}m{#1}}
\newcommand{\wimg}[1]{%
  \adjustbox{valign=c}{\includegraphics[width=24mm,height=4mm,keepaspectratio]{#1}}%
}
\setlength{\tabcolsep}{1pt}
\renewcommand{\arraystretch}{1.2}
\begin{table}[t]
\centering
\caption{\textsc{Comparison of on-chip training methods}}
\vspace{-4pt}
\begin{tabular}{l c c c C{20mm}}
\toprule
\makecell{\textbf{IMC}\\\textbf{works}} &
\makecell{\textbf{Non-ideality}\\\textbf{analysis}} &
\makecell{\textbf{Non-ideal}\\\textbf{behavior}} &
\makecell{\textbf{Linearity}\\\textbf{dependence}} &  
\makecell{\textbf{Weight} \\\textbf{distribution}} \\
\midrule
TCAD2019~\cite{8333741}             & \xmark & \xmark & \xmark & \wimg{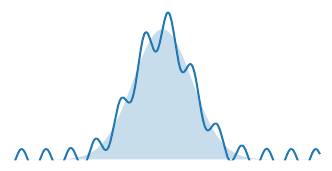} \\
TC2020~\cite{ankit2020panther}    & \xmark & \xmark & \xmark & \wimg{Images/Figure_00.pdf} \\
TETCI2021~\cite{9139434}            & \xmark & \xmark & \xmark & \wimg{Images/Figure_00.pdf} \\
TCASI2025~\cite{10753271}           & \xmark & \xmark & \xmark & \wimg{Images/Figure_00.pdf} \\
NCA2019~\cite{lim2019adaptive}       & \cmark & D2D & High & \wimg{Images/Figure_00.pdf} \\
TCASII2024~\cite{10233046}       & \cmark & C2C & High & \wimg{Images/Figure_00.pdf} \\
TCAD2021~\cite{9292971}             & \cmark & NL,D2D,C2C & High & \wimg{Images/Figure_00.pdf} \\
TCAD2022~\cite{9716051}             & \cmark & NL,D2D,C2C & High & \wimg{Images/Figure_00.pdf} \\
NC2023~\cite{rasch2023hardware}     & \cmark & NL,D2D,C2C & High & \wimg{Images/Figure_00.pdf} \\
NE2023~\cite{yi2023activity}     & \cmark & NL,D2D,C2C & High & \wimg{Images/Figure_00.pdf} \\
IVLSIJ2024~\cite{102206}     & \cmark & NL,D2D,C2C & High & \wimg{Images/Figure_00.pdf} \\
Sci.Adv2022~\cite{abm8537}            & \cmark & NL,D2D,C2C & High & \wimg{Images/Figure_00.pdf} \\
\textbf{This work}         & \cmark & NL,D2D,C2C & Low & \wimg{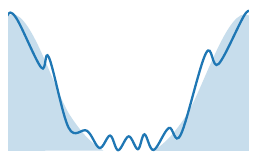} \\
\bottomrule
\end{tabular}
\vspace{-6pt}
\label{tab:Weight_Comparison}
\end{table}

\section{Background And Motivation}
\label{sec:section_2}
\subsection{eNVM Accelerators and On-Chip Training}
\label{sec:OnChip}
As illustrated in Fig.~\ref{fig:fig_2}(a), neural network training comprises four stages: forward propagation, backpropagation, gradient computation, and weight update. In eNVM-based accelerators, the implementation of these stages is described as follows.

\begin{figure}[t]
    \centering
    \includegraphics[width=1\linewidth]{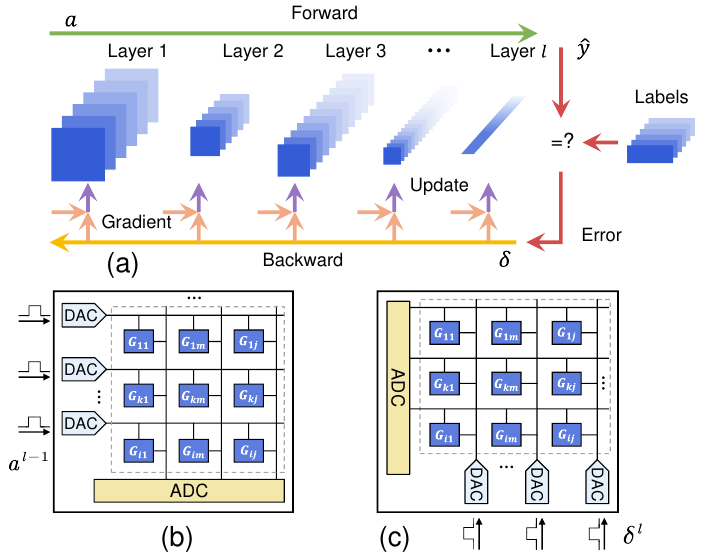}
    \vspace{-10pt}
    \caption{On-chip training implementation of IMC architecture: (a) neural network training workflow; (b) forward acceleration; (c) backward acceleration.}
    \vspace{-10pt}
    \label{fig:fig_1} 
\end{figure}

\textit{Forward Propagation:}
In forward propagation, the computation of the $l$-th layer can be expressed as:
\begin{equation}
a^l = \sigma(z^l) = \sigma(W^l \ast a^{l-1})
\label{eq:Eq_FP}
\end{equation}
where $\sigma(\cdot)$ denotes the activation function, $W^l$ represents the weights of the $l$-th layer, and $a^{l-1}$ is the activation from the preceding layer. To adapt to MVM-based accelerators, the Img2Col technique is typically employed to transform convolutions into general matrix multiplication (GEMM)~\cite{wan2022compute,yao2020fully}.
As shown in Fig.~\ref{fig:fig_1}(b), in eNVM accelerators, the weights are pre-programmed as device conductances. Input activations are converted to analog voltages by digital-to-analog converters (DACs) and then applied to the row lines of the eNVM array. According to Ohm's law and Kirchhoff's current law, input voltages and conductances perform the MVM computation within the array. The resulting currents are accumulated along the column lines and sensed and digitized by analog-to-digital converters (ADCs)~\cite{7551379,zhang2019design}. Finally, the error is computed from the final outputs and target values for backpropagation.

\textit{Error Backpropagation:}
During the backpropagation phase, the error is propagated from the output layer toward the input layer. The error of the $(l-1)$-th layer is given by:
\begin{equation}
\delta^{l-1} = \delta^l \ast rot180(W^l) \odot \sigma'(z^{l-1}) 
\label{eq:Eq_BP}
\end{equation}
where $\delta^l$ denotes the error of the $l$-th layer, $rot180(W^l)$ represents the $180^\circ$ rotation of the convolution kernel in the spatial domain, and $\odot$ denotes element-wise multiplication. After Img2Col expansion, this spatial rotation is mathematically equivalent to the transpose of the conductance matrix. As illustrated in Fig.~\ref{fig:fig_1}(c), in eNVM accelerators, the error is converted to analog voltage signals by DACs and applied to the column lines of the eNVM array. Subsequently, M$^{\mathrm{T}}$VM operations are performed within the eNVM array, and the analog computation results are measured by ADCs on the row lines~\cite{sebastian2020memory,ankit2020panther}. 
It should be noted that, unlike forward propagation, error backpropagation employs the transpose of the original weight matrix. Consequently, accelerating backpropagation with eNVM arrays requires additional hardware support to reverse the dataflow direction.

\textit{Weight Gradient Computation and Parameter Update:}
After obtaining the errors of each layer through backpropagation, the weight gradient of the $l$-th layer can be expressed as:
\begin{equation}
\frac{\partial \mathcal{L}}{\partial W^{l}} = \delta^l \ast \operatorname{rot180}(a^{l-1})
\label{eq:Eq_WG}
\end{equation}
where $\mathcal{L}$ denotes the loss function. 
After the weight gradients are obtained, the parameters are iteratively updated to minimize the loss function. The update rule is given by:
\begin{equation}
W^{l}_{\mathrm{new}} = W^{l}_{\mathrm{old}} - \alpha \cdot \frac{\partial \mathcal{L}}{\partial W^{l}}
\label{eq:Eq_WU}
\end{equation}
where $\alpha$ denotes the learning rate.

In eNVM-based accelerators, there are two implementation approaches for gradient computation and weight update~\cite{sebastian2020memory}. The first approach is the in-situ outer-product update method~\cite{sebastian2020memory,ankit2020panther,choi2021neural}. In this approach, the error $\delta^l$ and activation $a^{l-1}$ are encoded as programming pulses and applied to the row lines and column lines of the crossbar array, respectively. Through the coincidence of these pulses at the cross-points, the device conductance is directly modulated, completing the weight update within the array.
The second approach involves explicit gradient computation followed by weight writing~\cite{9292971,sebastian2020memory}. 
In this scheme, the weight gradients are first computed using peripheral circuits to generate the update $\Delta W$. Subsequently, the conductance of the eNVM cells is adjusted according to the magnitude and polarity of $\Delta W$, thereby completing the parameter update~\cite{choi2021neural,sebastian2020memory}.

\subsection{Non-ideal Weight Update}
\begin{figure}[t]
    \centering
    \includegraphics[width=0.99\linewidth]{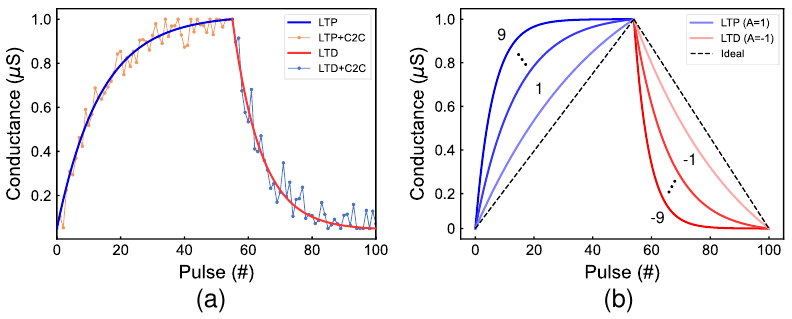}
    \vspace{-15pt}
    \caption{Non-ideal characteristics on eNVM devices. (a) Nonlinearity and C2C variation~\cite{10323793}. (b) Different degrees of nonlinearity~\cite{9292971}.}
    \vspace{-6pt}
    \label{fig:fig_2}
\end{figure}

As shown in Fig.~\ref{fig:fig_2}(a), when positive voltage pulses are applied to an eNVM device, its conductance gradually increases toward a maximum value, a process known as long-term potentiation (LTP). Conversely, applying negative voltage pulses causes the conductance to gradually decrease toward a minimum value, referred to as long-term depression (LTD)~\cite{9292971}. Ideally, the change in conductance during LTP and LTD should exhibit a linear relationship with the number of write pulses. However, in practical studies~\cite{10323793,9292971}, eNVM devices often deviate significantly from this ideal behavior. Specifically, the conductance changes rapidly during the initial stages of both LTP and LTD and then gradually saturates, resulting in pronounced nonlinear characteristics.

The nonlinear and asymmetric behavior can be modeled by the following Eqs.~(\ref{eq:Eq_1})–(\ref{eq:Eq_3})~\cite{9292971}, where the conductance change exhibits a nonlinear dependence on the pulse count $P$:
\begin{equation}
G_{\text{LTP}} = B \left( 1 - e^{-P/A} \right) + G_{\text{min}}
\label{eq:Eq_1}
\end{equation}
\begin{equation}
G_{\text{LTD}} = -B \left( 1 - e^{(P-P_{\text{max}})/A} \right) + G_{\text{max}}
\label{eq:Eq_2}
\end{equation}
\begin{equation}
B = (G_{\text{max}} - G_{\text{min}}) / (1 - e^{-P_{\text{max}} / A})
\label{eq:Eq_3}
\end{equation}
Here, $G_{\text{max}}$ and $G_{\text{min}}$ represent the maximum and minimum conductance values attainable by the device, respectively, and $P_{\text{max}}$ denotes the maximum pulse number. As illustrated in Fig.~\ref{fig:fig_2}(b), the parameter $A$ determines the degree of nonlinearity in the conductance update: a smaller value of $A$ results in conductance evolution closer to linear behavior, whereas a larger value of $A$ leads to more pronounced nonlinearity. The parameter $B$ is a normalization factor determined by $G_{\text{max}}$, $G_{\text{min}}$, and $P_{\text{max}}$. $G_{\mathrm{LTP}}$ and $G_{\mathrm{LTD}}$ correspond to the conductance values in the LTP and LTD phases, respectively.

In addition to nonlinearity, eNVM devices face two critical challenges: C2C and D2D variations, as illustrated in Fig.~\ref{fig:fig_2}(a). C2C variation primarily originates from dynamic changes in the internal structure and material properties of a device across different write/erase cycles, leading to inconsistent electrical characteristics~\cite{9775004}. D2D variation arises from differences in fabrication processes, material composition, and microstructure, resulting in significant deviations in electrical characteristics among different devices~\cite{10070583}. During on-chip training of neural networks, these non-idealities can cause weight updates to deviate from target values, reducing the precision of weight adjustment. Previous studies~\cite{9406197,9292971} based on the VGG8 network and the CIFAR10 dataset have shown that training accuracy decreases significantly when devices exhibit severe asymmetry, with an accuracy degradation of up to 25.0\%. This observation highlights the necessity of mitigating the nonideal characteristics of eNVM devices.

\begin{figure}[t]
    \centering
    \includegraphics[width=1\linewidth]{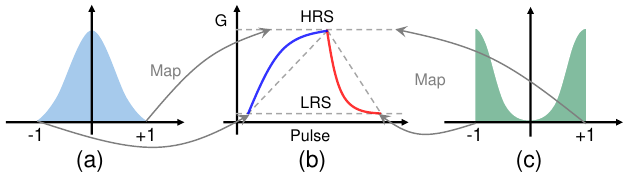}
    \vspace{-15pt}
    \caption{(a) Weight distribution from conventional training; (b) nonlinear conductance modulation; (c) weight distribution after NAT training.}
    \vspace{-8pt}
    \label{fig:fig_3}
\end{figure}

\subsection{Strategies for Mitigating Non-Idealities in eNVM Devices}
\label{sec:Strategies}
In conventional neural networks, the weight distribution is typically approximately Gaussian with a mean close to zero, resulting in most weights clustering around zero, as illustrated in Fig.~\ref{fig:fig_3}(a). When these weights are mapped onto eNVM devices, they primarily occupy the intermediate analog conductance states. However, this intermediate region is precisely where device non-idealities are most pronounced, making it highly susceptible to NL, C2C, and D2D variations. Consequently, conductance values become unstable and difficult to control precisely, which constitutes a key factor underlying the degradation of on-chip training accuracy~\cite{9406197,9292971}.

In contrast, when the device operates in the high-resistance state (HRS) or low-resistance state (LRS), it exhibits the following advantages. First, at HRS and LRS, the conductive channel is either fully formed or completely ruptured; consequently, even when subjected to physical processes such as thermal fluctuations or defect migration~\cite{10070583}, the device is unlikely to exhibit significant conductance drift, yielding optimal retention characteristics~\cite{Li2024}. Second, within the entire nonlinear conductance tuning range, only at HRS and LRS is the weight-to-conductance mapping deterministic, with mapping deviation approaching zero. Third, the sensitivity of HRS and LRS to programming pulses is substantially lower than that of intermediate conductance states, mitigating the impact of programming errors on weight update precision. Finally, when weights reside at HRS or LRS, fine‑grained conductance adjustment is unnecessary, which reduces the number of write iterations and lowers write energy consumption~\cite{article,abm8537}. 

Motivated by the above analysis of neural network weight distributions and the physical characteristics of eNVM devices, we propose the NAT algorithm. As shown in Fig.~\ref{fig:fig_3}(c), the core idea of NAT is to steer weight distribution toward the HRS and LRS regions during training, thus fully leveraging the advantages of these two states. To achieve this, we introduce a strong regularization constraint at the algorithmic level that gradually drives the weights toward HRS or LRS during optimization iterations. This strategy exploits the superior conductance retention of HRS and LRS while circumventing the highly nonlinear intermediate conductance region, effectively suppressing the impact of write noise on training accuracy from the device physics perspective. Overall, the approach embodies an algorithm-device co-design philosophy, enhancing practical deployment feasibility by aligning weight distribution with inherent hardware characteristics.

\begin{figure}[!t]
    \centering
    \includegraphics[width=1\linewidth]{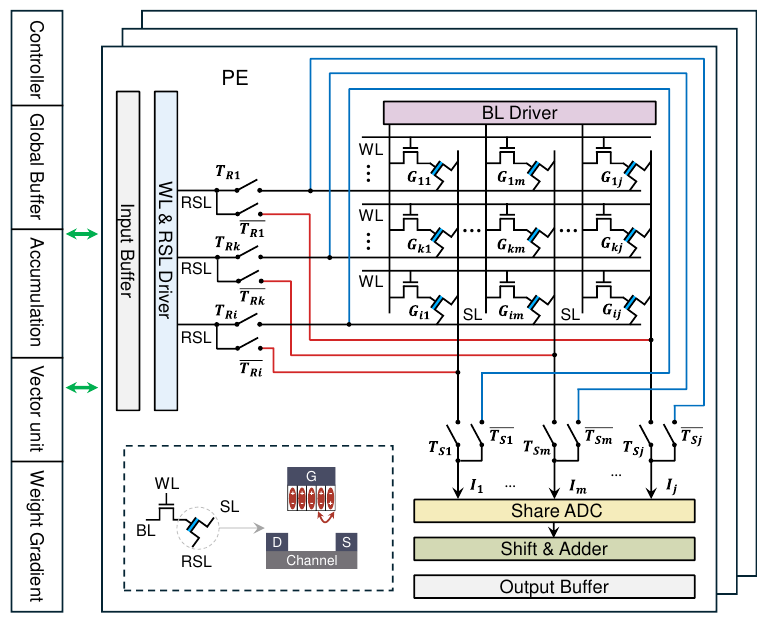}
    \vspace{-12pt}
    \caption{NOVA on-chip training architecture. The dashed box illustrates the cross-sectional structure of the FeFET device.}
    \vspace{-6pt}
    \label{fig:fig_4} 
\end{figure}

\section{Methodology}
\label{sec:section_3}
This section first provides an overview of the proposed NOVA accelerator architecture. Subsequently, using our developed FeFET device as a representative eNVM, it presents a bottom-up hierarchical description, including the fabrication and behavioral modeling of the FeFET device, the design and implementation of the NAT algorithmic, and the software framework of the on-chip training simulation.

\subsection{NOVA Architecture}
\subsubsection{Architecture Overview}
In this section, we present NOVA, a cross-layer design architecture tailored for on-chip training of neural networks.  As illustrated in Fig.~\ref{fig:fig_4}, NOVA primarily comprises multiple processing engines (PEs), a global buffer, a weight-gradient unit, a vector unit, and an accumulation unit. The global buffer stores training data and intermediate results generated during computation. The weight-gradient unit is dedicated to weight-gradient computation during on-chip training. The vector unit performs non-array operations, such as activation and classification, while the accumulation unit aggregates intermediate results from multiple PEs to produce the final outputs. Each tile is interconnected through an H-tree routing network. Throughout the on-chip training process, the controller dynamically schedules data paths and computational tasks according to different training stages to improve computational efficiency and hardware resource utilization.

In the NOVA architecture, each PE integrates an FeFET array, input/readout circuitry, and configurable dataflow control logic. As shown in the inset within the dashed box in Fig.~\ref{fig:fig_4}, the FeFET is a three-terminal device. In an array composed of 1T1F (one-transistor-one-FeFET) cells, the read operation primarily consists of three steps. First, a selection signal is applied via the wordline (WL) to activate the access transistor and establish a conductive path. Second, a read voltage is applied to the FeFET device through the read select line (RSL). Finally, the channel current of the device is collected via the bitline (BL) and transmitted to the sensing circuitry. During computational acceleration, the access transistors in the crossbar array are activated via the WL, and the read voltages are applied to the FeFET devices through the RSL. Based on physical laws, analog MVM and M$^{\mathrm{T}}$VM operations are performed within the crossbar array. The resulting currents are accumulated on the source lines (SL) and then sampled and digitized by shared ADCs, yielding the multiply-accumulate results for forward and backward propagation.

\subsubsection{On-Chip Training Implementation}
\begin{figure}[t]
    \centering
    \includegraphics[width=1\linewidth]{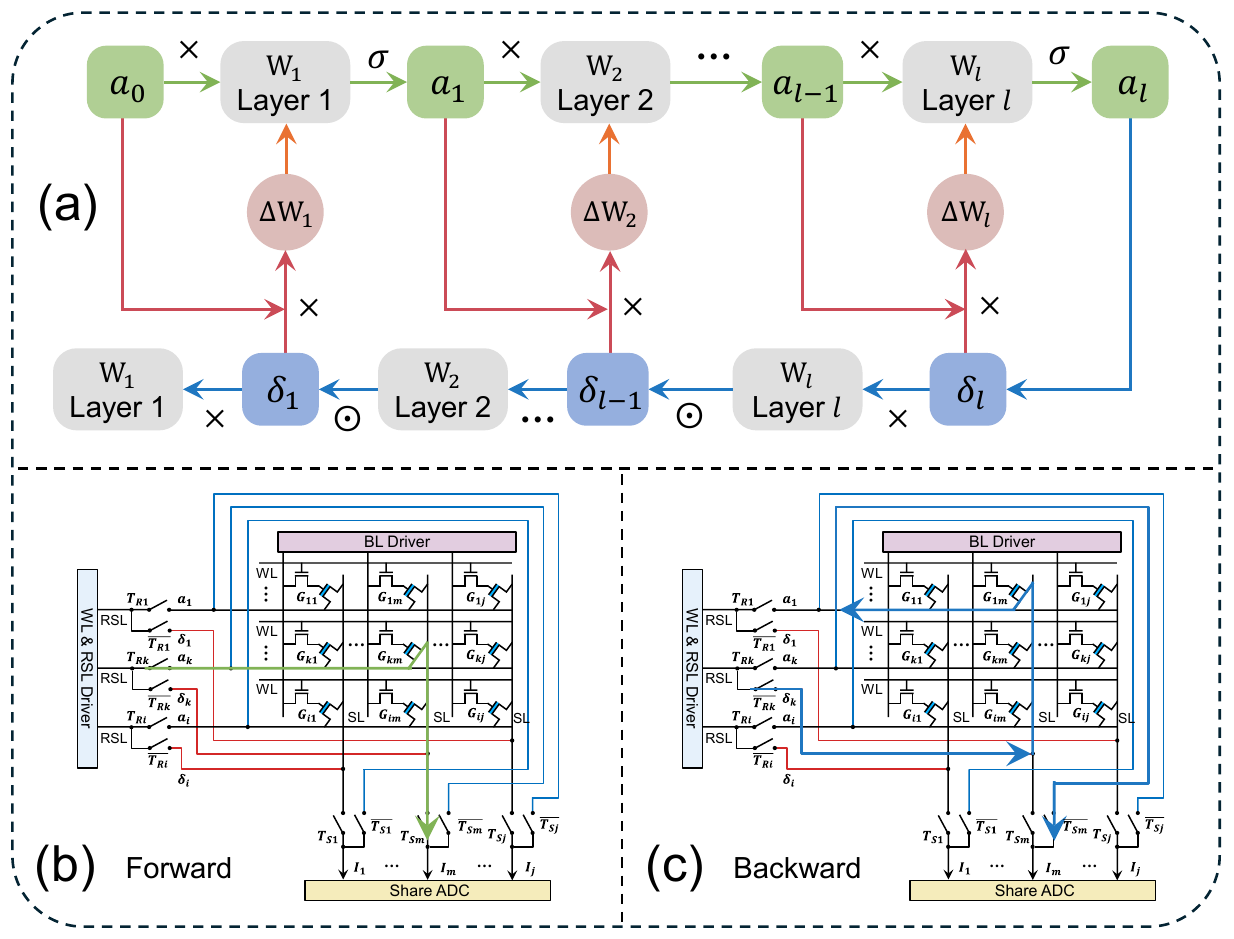}
    \vspace{-12pt}
    \caption{Dataflow and computational mechanisms in NOVA. (a) Neural network training workflow; (b) Forward acceleration; (c) Backward acceleration.}
    \vspace{-8pt}
    \label{fig:fig_5}
\end{figure}

To support flexible dataflow scheduling during on-chip training, NOVA integrates configurable switch circuits in each PE for dynamically switching between forward and backward tasks. These switch circuits are implemented using transmission gates, resulting in minimal area and power overhead. By configuring the switch states, this design not only enables the transposition of the conductance matrix but also allows peripheral hardware resources to be shared across different computation phases, thereby enhancing overall hardware utilization and energy efficiency. In conjunction with the neural network training workflow illustrated in Fig.~\ref{fig:fig_5}(a), NOVA's specific support mechanisms for each training phase are detailed as follows.

During forward propagation, when switches $S_{R1}$, $S_{Rk}$, $S_{Ri}$, $S_{S1}$, $S_{Sm}$, and $S_{Sj}$ are closed, the system performs the forward MVM operation described by Eq.~(\ref{eq:Eq_FP}). The input signals are injected through the row lines, while the resulting currents are accumulated and read out along the column lines. The corresponding dataflow is illustrated in Fig.~\ref{fig:fig_5}(b). During the backpropagation, the connection relationship between the row lines and column lines of the array is reversed when the complementary switches $\overline{S_{R1}}$, $\overline{S_{Rk}}$, $\overline{S_{Ri}}$, $\overline{S_{S1}}$, $\overline{S_{Sm}}$, and $\overline{S_{Sj}}$ are closed. Consequently, the system executes backward M$^{\mathrm{T}}$VM computation corresponding to Eq.~(\ref{eq:Eq_BP}).
In this configuration, the input signals are injected through the column lines, while the resulting currents are accumulated and read out along the row lines. The corresponding dataflow is illustrated in Fig.~\ref{fig:fig_5}(c). 
During gradient computation and parameter update, the weight gradient unit first computes the weight gradients according to Eq.~(\ref{eq:Eq_WG}) and derives the corresponding parameter update $\Delta W$ using Eq.~(\ref{eq:Eq_WU}). Based on the signs and magnitudes of $\Delta W$, the driver circuitry applies programming pulses of the corresponding polarities and widths to the gates of the target FeFET devices. These pulses alter the polarization states of the ferroelectric gate dielectric, thereby modulating the channel conductance and completing the conductance updates.

\begin{figure*}[!t]
    \centering
    \includegraphics[width=1\linewidth]{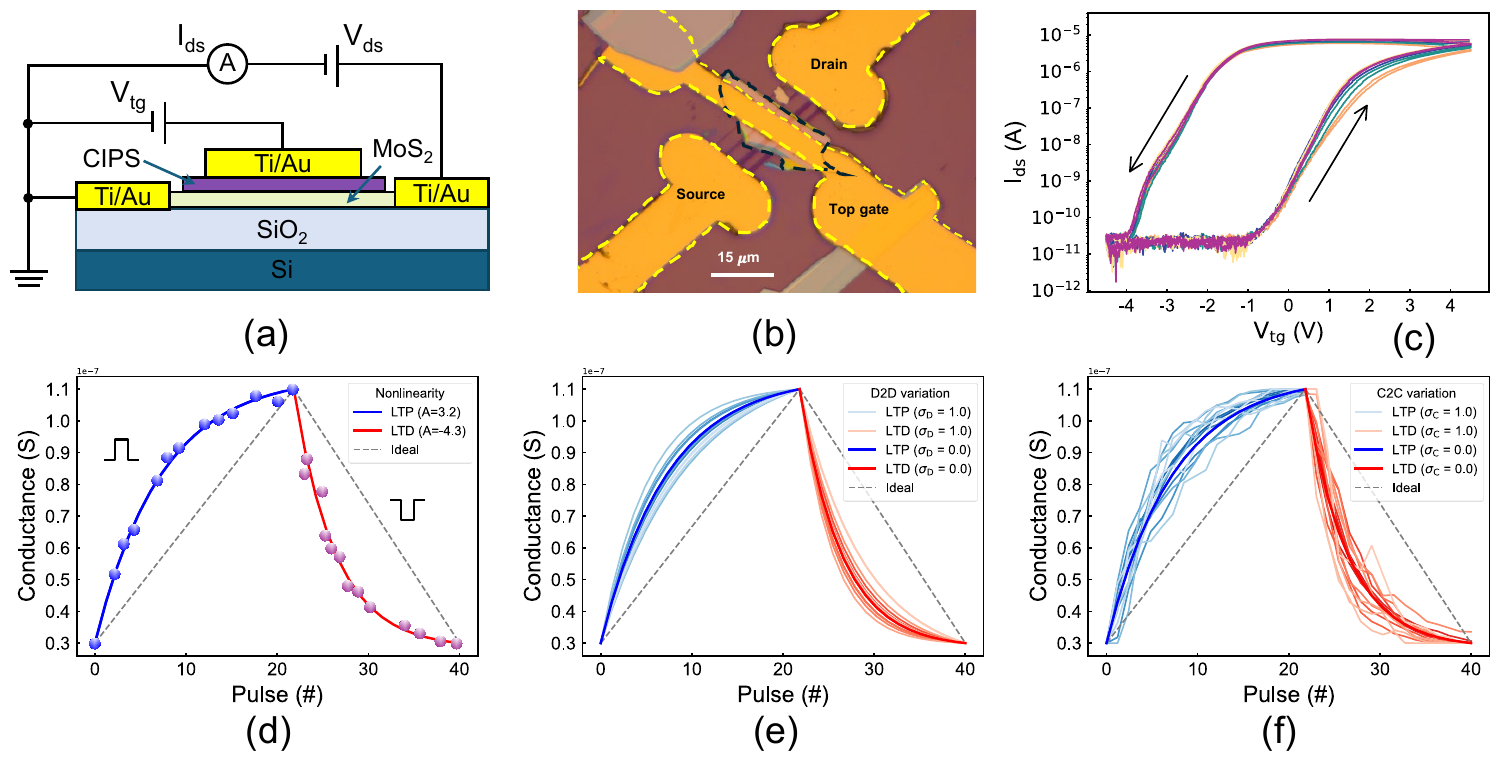}
    \vspace{-12pt}
    \caption{Fabrication of a CuInP$_2$S$_6$/MoS$_2$-based FeFET and modeling of its non-idealities: (a) schematic of the device structure; (b) optical micrograph; (c) measured \(I_{\mathrm{DS}}{-}V_{\mathrm{G}}\) transfer characteristics with the \(\mathrm{SiO}_{2}\) back gate floating and \(V_{\mathrm{DS}}=1.0~\mathrm{V}\); (d) nonlinear conductance update model fitted to measurement data; (e) D2D variability model; (f) C2C variability model.}
    \vspace{-4pt}
    \label{fig:fig_6}
\end{figure*}

\subsection{Device Fabrication and Behavioral Modeling}
\label{sec:Modeling}
\subsubsection{Heterostructure preparation and device fabrication}
In this section, we present the fabrication of a FeFET device based on 2D ferroelectric CuInP$_2$S$_6$ (CIPS) and semiconducting MoS$_2$. Fig.~\ref{fig:fig_6}(a) depicts the schematic structure of the FeFET device. The flakes of MoS$_2$ and CIPS were mechanically exfoliated from single-crystal bulk materials. The heterostructure was assembled using a previously reported dry-transfer technique~\cite{gao2024realization}. First, multilayer MoS$_2$ flakes were exfoliated onto a SiO$_2$/Si substrate, and CIPS flakes were exfoliated onto a transparent polydimethylsiloxane (PDMS) film. Then, using a 2D transfer platform inside a nitrogen-filled glove box, the PDMS film carrying the CIPS flake was aligned with the MoS$_2$ flake, and the CIPS was transferred to MoS$_2$. Finally, standard photolithography was employed to pattern  Ti/Au (10 nm/50 nm) electrodes on the heterostructure, followed by metal deposition via electron-beam evaporation and a lift-off process. The optical image of the FeFET device is shown in Fig.~\ref{fig:fig_6}(b). Under the sequence of top-gate voltage pulses, progressive reversal of the ferroelectric domains in the ferroelectric CIPS enables modulation of multiple conductance states in the MoS$_2$ channel. The transfer characteristics ($I_{\text{ds}}$–$V_{\text{g}}$) measured under the top gate (when the drain–source voltage $V_{\text{ds}} = 1.0$ V) are presented in Fig.~\ref{fig:fig_6}(c). In terms of nonvolatile memory performance, the device exhibits a large memory window and an on/off current ratio as high as $10^6$.

\subsubsection{Nonlinear Behavior Model}
To model the multi-level conductance modulation behavior of FeFET devices, we fitted the conductance update characteristics using experimental data measured at $V_{\text{g}} = 0~\text{V}$ and $V_{\text{ds}} = 0.1~\text{V}$ according to equations~(\ref{eq:Eq_1})--(\ref{eq:Eq_3}). The optimal fitting parameters were obtained as $A_{\text{LTP}} = 3.2$ and $A_{\text{LTD}} = -4.3$. Based on these parameters, the final conductance update expressions are given as follows:
\begin{equation}
G_{\text{LTP}} = \left( \frac{G_{\text{max}} - G_{\text{min}}}{1 - e^{-P_{\text{max}} / A_{\mathrm{LTP}}}} \right) \left(1 - e^{-P / A_{\mathrm{LTP}}}\right) + G_{\text{min}}
\end{equation}
\begin{equation}
G_{\text{LTD}} = -\left( \frac{G_{\text{max}} - G_{\text{min}}}{1 - e^{-P_{\text{max}} / A_{\mathrm{LTD}}}} \right) \left(1 - e^{(P - P_{\text{max}}) / A_{\mathrm{LTD}}}\right) + G_{\text{max}}
\end{equation}
where $G_{\mathrm{LTP}}$ and $G_{\mathrm{LTD}}$ denote the conductance in the LTP and LTD phases, respectively, and $P$ represents the number of programming pulses. The update behavior for conductance decrease can be expressed as $P' = P_{\max}-P$. The fitting results are shown in Fig.~\ref{fig:fig_6}(d).

\subsubsection{D2D Behavior Model}
In constructing the D2D variation model, we introduce Gaussian perturbations into the nonlinear parameters fitted for the LTP and LTD phases to capture the physical variations caused by fabrication process fluctuations. Specifically, the nonlinear parameters of each individual device ($\widehat{A}_{\mathrm{LTP}}$ and $\widehat{A}_{\mathrm{LTD}}$) are sampled as follows:
\begin{equation}
\begin{aligned}
\widehat{A}_{\mathrm{LTP}} &= A_{\text{LTP}} + \mathcal{N}(0, \sigma_{\text{D}}^2) \\
\widehat{A}_{\mathrm{LTD}} &= A_{\text{LTD}} + \mathcal{N}(0, \sigma_{\text{D}}^2)
\end{aligned}
\end{equation}
where $\sigma_{\text{D}}$ denotes the standard deviation of inter-device nonlinear perturbations, and $\mathcal{N}(0, \sigma_{\text{D}}^2)$ denotes a zero-mean Gaussian random variable with variance $\sigma_{\text{D}}^2$. As shown in Fig.~\ref{fig:fig_6}(e), the curve colors from dark to light correspond to ten D2D variation intensity levels with $\sigma_{\text{D}} \in \{0, 0.1, \dots, 1.0\}$. This model characterizes the conductance update trajectories across different devices.

\subsubsection{C2C Behavior Model}
To model C2C variations in eNVM devices, we introduce a multiplicative noise model, in which the noise is applied as a scaling factor to the weight update amount, yielding a perturbation amplitude that scales linearly with $\Delta W$. The model is expressed as follows:
\begin{equation}
W_{\text{new}} = W_{\text{old}} + \Delta W \cdot \left(1 + \mathcal{N}(0, \sigma_{\text{C}}^2)\right)
\end{equation}
where $W_{\text{old}}$ and $W_{\text{new}}$ denote the weights before and after the update, respectively. $\Delta W$ represents the weight update amount, and $\mathcal{N}(0, \sigma_{\text{C}}^2)$ denotes zero-mean Gaussian noise with variance $\sigma_{\text{C}}^2$. Compared with the additive noise model with a fixed noise amplitude used in~\cite{9292971}, this multiplicative noise model better reflects the physical mechanism of device programming, in which the perturbation magnitude is positively correlated with the conductance update amplitude. In addition, to strictly enforce the weight boundary constraint, a quantization-aware clipping operation, $\text{Clamp}(\cdot, \text{bits}_W)$, is applied to the weights after each update. As shown in Fig.~\ref{fig:fig_6}(f), the curve colors from dark to light correspond to ten C2C variation intensity levels with $\sigma_{\text{C}} \in \{0, 0.1, \dots, 1.0\}$. This model captures the conductance evolution behavior across different programming cycles.

\subsection{Training Algorithm Tailored for Non-ideal eNVM}
Based on the analysis in Section \ref{sec:Strategies}, we propose a training algorithm named NAT that mitigates the adverse effects caused by non-idealities of the eNVM device. 
In this work, we adopt the WAGE~\cite{wu2018training} training method, which quantizes weights, activations, gradients, and errors to 8-bit precision. Through quantization‑aware training (QAT), the computational tensors are quantized to low precision, ensuring that the eNVM accelerator can operate stably within its limited conductance states. Furthermore, to match the dynamic range of the hardware, the values of weights and activations are restricted to $[-1, +1]$. To drive weights to converge toward the boundary values ($\pm 1$) during training, we introduce a regularization method that promotes polarization of the weight distribution by driving the magnitudes of weights toward unity. Specifically, we incorporate the regularization term over all learnable weights (i.e., parameters of convolutional and fully connected layers) and jointly optimize it with the task loss:
\begin{equation}
\mathcal{L}_{\text{total}} = \mathcal{L}_{\text{task}} + \lambda(t) \cdot \mathcal{L}_{\text{mag}}, \quad
\mathcal{L}_{\text{mag}} = \sum_{j} \big|\, |w_j| - 1 \,\big|_p
\vspace{-3pt}
\end{equation}
where $\mathcal{L}_{\text{task}}$ denotes the task-specific loss (e.g., cross-entropy), and $\mathcal{L}_{\text{mag}}$ measures the $p$-norm deviation of the weight magnitudes from unity, with $p=1$ used by default in this work. The regularization coefficient $\lambda(t)$ follows a piecewise linear warm-up schedule to avoid imposing overly aggressive constraints on the weights during the early training phase, thereby mitigating the risk of gradient anomalies and ensuring smooth convergence toward the optimal trajectory.

To optimize the parameters of the model through gradient descent, the partial derivative of the total loss function $\mathcal{L}_{\text{total}}$ with respect to weight $w_i$ must be computed. By the linearity of differentiation, this gradient decomposes into the sum of the gradient of the task loss and the scaled gradient of the regularization term. The gradient of the task loss, $\frac{\partial \mathcal{L}_{\text{task}}}{\partial w_i}$, is obtained through standard backpropagation, while the gradient of the regularization term requires separate derivation. The gradient of the regularization term with respect to $w_i$ is:
\begin{equation}
\frac{\partial \mathcal{L}_{\text{mag}}}{\partial w_i} = \frac{\partial}{\partial w_i} \big|\, |w_i| - 1 \,\big|
\end{equation}
Since the absolute value function is non-differentiable at zero, we employ the subgradient for handling this case. Specifically, the subgradient of $|x|$ is the sign function $\mathrm{sgn}(x)$ with $\mathrm{sgn}(0)=0$. Applying the chain rule, the gradient of the regularization term becomes $\mathrm{sgn}(|w_i| - 1) \cdot \mathrm{sgn}(w_i)$. Consequently, the gradient of the total loss with respect to $w_i$ is expressed as:
\begin{equation}
\frac{\partial \mathcal{L}_{\text{total}}}{\partial w_i}
= \frac{\partial \mathcal{L}_{\text{task}}}{\partial w_i}
+ \lambda(t)\cdot\mathrm{sgn}(|w_i| - 1) \cdot \mathrm{sgn}(w_i)
\end{equation}
where the second term imposes a magnitude constraint whenever $|w_i| \neq 1$, actively pushing the weight toward the target magnitude of 1 and shaping the weight distribution.

With the introduction of the NAT regularization term, $\mathcal{L}_{\mathrm{total}}$ serves as the overall optimization objective. Specifically, the task loss $\mathcal{L}_{\mathrm{task}}$ optimizes the feature learning capability and task performance of the network, while the regularization term $\mathcal{L}_{\mathrm{mag}}$ drives weights toward the boundary regions; the coefficient $\lambda$ controls the regularization strength. During training, the interplay between the task loss gradient and the regularization gradient imposes a soft constraint on the weight distribution. When migrating a particular weight toward the boundary regions would significantly increase the task loss, the gradient from $\mathcal{L}_{\mathrm{task}}$ counteracts or suppresses the regularization-driven push. Conversely, for weights that are insensitive to the task loss, the regularization term steers them toward the boundary regions. This soft constraint adaptively modulates the weight density within the conductance window, increasing the proportion of weights in the boundary regions while still permitting task-critical weights to remain in the intermediate conductance region, thereby avoiding rigidly confining weights to a few discrete conductance states. Ultimately, this joint gradient optimization achieves a balance between maintaining model accuracy and reshaping the weight distribution.

The proposed regularization method constructs an ``attractor basin'' in the parameter space centered at $|w| = 1$, guiding the weights to progressively converge toward $\pm 1$ while minimizing the classification error. Such a bipolar weight distribution offers multiple advantages. First, the positive and negative weight boundaries correspond to the physical boundaries of eNVM devices (i.e., HRS and LRS). These two conductance states feature well-defined boundaries and enhanced stability. Second, by increasing the proportion of weights in the boundary regions while reducing the weight density in the intermediate nonlinear conductance region, NAT maps more weights to relatively stable and approximately linear regions, effectively reducing mapping deviations and update errors. 
Overall, this algorithm--device co-design approach effectively leverages to the physical characteristics of eNVM devices while preserving model trainability.

Among existing regularization techniques, L1 (Lasso) regularization achieves sparsity through the penalty term $\sum_{j} |w_j|$ to facilitate feature selection. In contrast, L2 (Ridge) regularization compresses weights to a small range via the penalty term $\sum_{j} w_j^2$ to prevent overfitting~\cite{obi2023review}. However, both L1 and L2 tend to drive weights toward zero, causing numerous weights to fall into regions of pronounced hardware non-idealities in eNVM accelerators, increasing computational errors. In contrast, the proposed NAT method has explicit physical orientation, mitigating device non-idealities by reshaping the weight distribution to hardware-friendly regions. This establishes NAT's unique position in the regularization spectrum as a method with physical awareness capability. Additionally, NAT exhibits computational complexity comparable to L1/L2 regularization and can be efficiently implemented through vector operations in deep learning frameworks, introducing negligible additional computational overhead.

\begin{figure}[!t]
    \centering
    \includegraphics[width=0.99\linewidth]{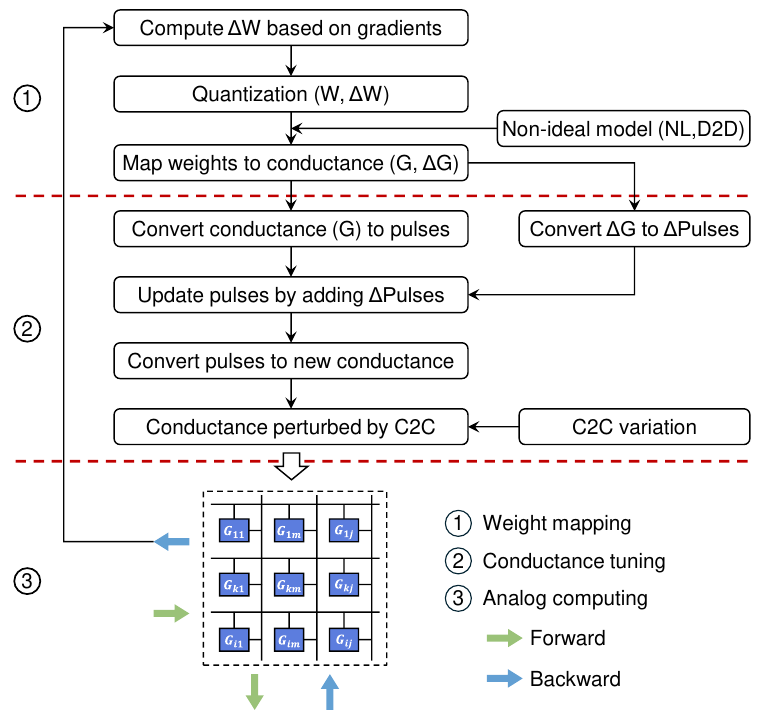}
    \vspace{-18pt}
    \caption{Workflow of a single iteration during on-chip training.}
    \vspace{-8pt}
    \label{fig:fig_7}
\end{figure}

\subsection{Simulation Framework for On-Chip Training with eNVM}
To simulate the on-chip training process of eNVM accelerators, we developed a complete software stack. As illustrated in Fig.~\ref{fig:fig_7}, the proposed on-chip training framework comprises three key phases during each training iteration: weight mapping, conductance tuning, and analog computation.

\textit{Weight Mapping:}
During the weight mapping phase, the ideal weight update $\Delta W$ is first computed based on the gradient. A uniform scaling factor is then applied to adjust the numerical range of $\Delta W$ to match the programmable conductance window of the eNVM device. Subsequently, a stochastic rounding strategy~\cite{wu2018training} is employed to quantize the scaled weight update, mapping continuous values to a finite set of conductance states. Finally, following the methodology described in Section \ref{sec:Modeling}, a non-ideal device behavior model incorporating NL and D2D variations is constructed, and the quantized weights and weight updates are converted into actual eNVM's conductance values $G$ and $\Delta G$.

\textit{Conductance Tuning:}
In the conductance tuning phase, conductance updates are implemented through a pulse accumulation mechanism. Specifically, the current conductance value is first converted into a pulse representation, and the conductance increment $\Delta G$ is translated into a pulse increment $\Delta\text{Pulses}$. After the conductance update is performed by pulse accumulation, the updated pulse value is remapped to a new conductance state. Finally, C2C variations are introduced into the updated conductance values to emulate the stochasticity inherent in the actual write process.

\textit{Analog Computation:}
During the analog computation phase, the conductance array, after nonlinear mapping and C2C perturbation, is employed to perform analog MVM and M$^{\mathrm{T}}$VM operations. By configuring the switching circuitry around the array, the dataflow direction can be switched, enabling forward and backward propagation computations, respectively. Finally, the output results from the crossbar array are read and decoded to complete the neural network computation.

\section{Evaluation}
\label{sec:section_4}
\subsection{Experimental Setup}

\setlength{\tabcolsep}{18pt} 
\renewcommand{\arraystretch}{1.68}
\begin{table}[t]
\centering
\caption{\textsc{Hyperparameter Settings for Model Training}}
\vspace{-4pt}
\begin{tabular}{ccc}
\toprule
\textbf{Category} & \textbf{Hyperparameter} & \textbf{Value} \\
\hline
\multirow{3}{*}{\makecell{Optimization \\ \& regularization}} & Momentum & 0.9 \\
\cline{2-3}
 & NAT norm order & 1 \\
\cline{2-3}
 & {$T_{\text{warmup}}$} & 5 \\
\hline
\multirow{3}{*}{Training setup} & Epochs & 300 \\
\cline{2-3}
 & Batch size & 128 \\
\cline{2-3}
 & Initial learning rate & 0.1 \\
\bottomrule
\end{tabular}
\label{tab:training_params}
\vspace{-6pt}
\end{table}

\subsubsection{Model Training}
In this work, model training was conducted using a PyTorch-based framework that integrated optimization strategies such as SGD with momentum, learning rate warm-up, and data augmentation. The framework also incorporated the proposed eNVM conductance update model and the NAT algorithm, enabling hardware-aware on-chip training under the combined effects of device non-idealities and algorithmic design. For evaluation, we selected three network architectures (VGG6, VGG11, and ResNet18) and three datasets (SVHN, CIFAR10, and CIFAR100) to conduct paired experiments on the proposed NOVA architecture, assessing the performance of the proposed method across different networks and classification tasks. The main training hyperparameters were summarized in Table~\ref{tab:training_params}.

\subsubsection{Hardware Evaluation}
To evaluate the hardware performance of the NOVA architecture, we developed a customized hardware evaluation framework based on the open-source simulators NeuroSim~\cite{9292971} and MNSim~\cite{10058114}. The accumulation units, buffers, and weight-gradient computation unit adopt the existing circuit models and evaluation methods provided by NeuroSim. The compute units for forward and backward propagation are implemented using 1T1F crossbar arrays constructed with CIPS/MoS$_2$ heterojunction-based FeFET devices. For the dataflow switching circuitry specific to NOVA, we employ an analytical model parameterized by the predictive technology model (PTM)~\cite{zhao2006ptm} and the cache access and cycle time (CACTI) model~\cite{muralimanohar2009cacti} to evaluate its hardware performance. The system clock frequency is set to 1$~\mathrm{GHz}$. The readout path is equipped with shared 6-bit ADCs to sample and digitize the results of forward and backward computations, balancing conversion precision and hardware overhead. The detailed hardware parameters are listed in Table~\ref{tab:hardware_params}.

\setlength{\tabcolsep}{14.3pt}
\renewcommand{\arraystretch}{1.68}
\begin{table}[t]
\centering
\caption{\textsc{Hardware Simulation Parameter Settings}}
\vspace{-4pt}
\begin{tabular}{ccc}
\toprule
\textbf{Component} & \textbf{Parameter} & \textbf{Value} \\
\hline
\multirow{3}{*}{FeFET device} & $A_{\mathrm{LTP}}/A_{\mathrm{LTD}}$ & 3.2 / -4.3 \\
\cline{2-3}
 & LRS/HRS & 9\,M$\Omega$/33\,M$\Omega$ \\
\cline{2-3}
 & Read/Write voltage & 0.5\,V/1\,V  \\
\hline
\multirow{2}{*}{Crossbar} 
 & Technology & 32\,nm \\
\cline{2-3}
 & Subarray size & 128$\times$128 \\
\hline
ADC & Resolution & 6~bit \\
\bottomrule
\end{tabular}
\label{tab:hardware_params}
\vspace{-6pt}
\end{table}

\subsection{Evaluation of the NAT Algorithm on Weight Distribution}
In this section, we evaluate the impact of the NAT algorithm on the weight distribution and identify the optimal range of the NAT regularization strength. We employ the CIPS FeFET device fabricated in this work and, based on the device behavioral model described in Section \ref{sec:Modeling}, conduct comparative experiments within a unified simulation framework to systematically assess the regulation effect of NAT.

\subsubsection{Evaluation of the Impact of NAT Algorithm on Weight Distribution}
\begin{figure}[t]
    \centering
    \includegraphics[width=0.95\linewidth]{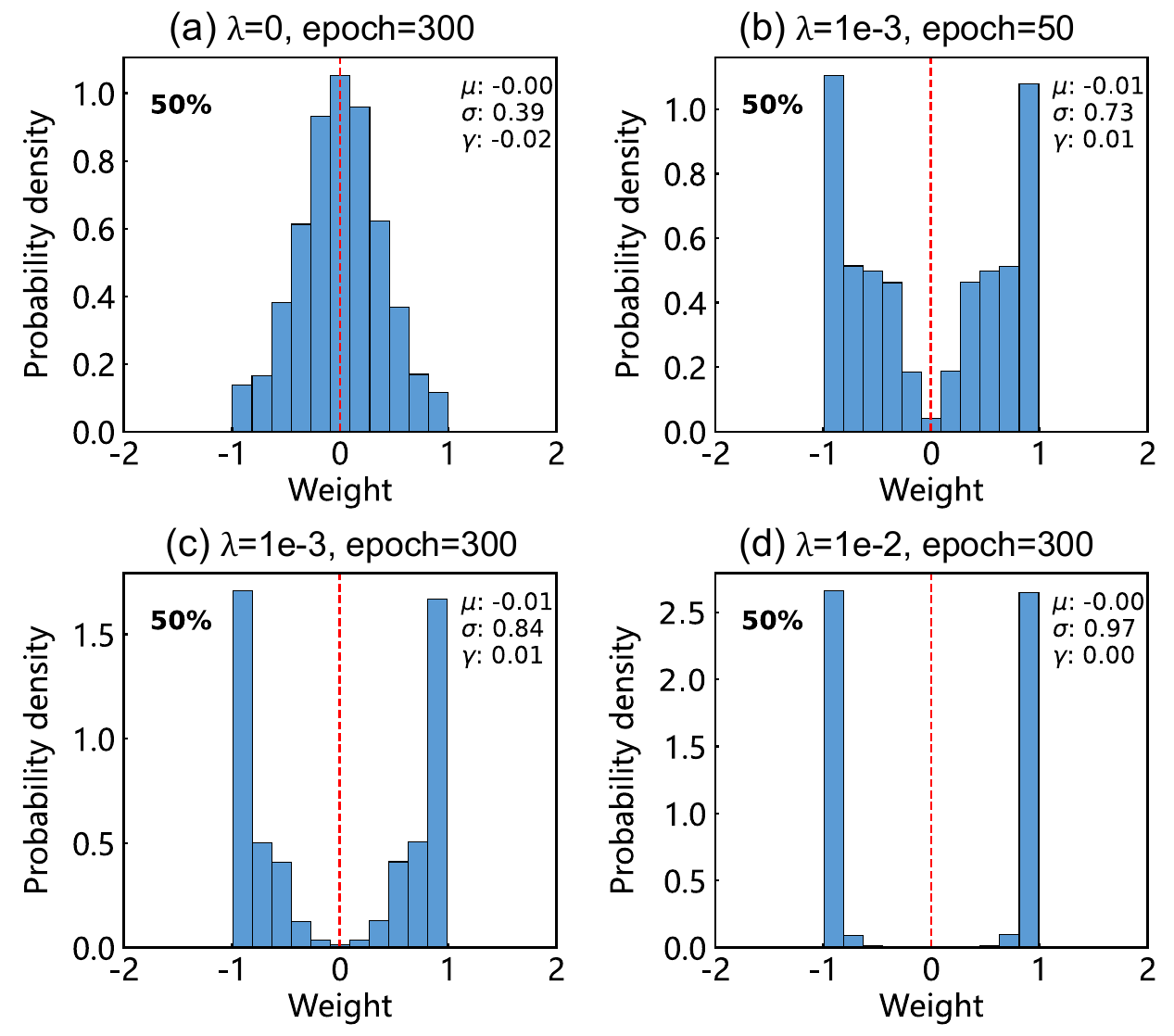}
    \vspace{-5pt}
    \caption{Evolution and comparison of the layer-6 weight distribution in VGG11: (a) conventional training yields a near-zero-mean unimodal distribution; (b), (c) with NAT, the distribution gradually converges toward a bimodal pattern at \(\pm 1\) by epochs 50 and 300, respectively; (d) with a regularization coefficient $\lambda = 1\mathrm{e}{-}2$, the bimodality becomes more pronounced. Here, \(\mu\) denotes the mean, \(\sigma\) the standard deviation, and \(\gamma\) the skewness.}
    \vspace{-6pt}
    \label{fig:fig_8}
\end{figure}

Fig.~\ref{fig:fig_8} illustrates the evolution of weight distributions in the sixth layer of VGG11 under different training strategies and hyperparameter settings. In conventional training without regularization (Fig.~\ref{fig:fig_8}(a)), the weights follow a near-zero-mean Gaussian-like unimodal distribution with a standard deviation of $\sigma = 0.45$ and skewness of $0.02$, with a large fraction of weights concentrated around zero. Upon introducing the NAT regularization term $\mathcal{L}_{\text{mag}} = \sum_i \big\lvert\, |w_i| - 1 \,\big\rvert_p$, the penalty on weight magnitudes in the loss function drives the weights away from zero toward the $\pm 1$ boundaries. An incipient bimodal structure is already visible in the early training stage (Fig.~\ref{fig:fig_8}(b), $\lambda = 1\mathrm{e}{-}3$, epoch = 50), with the standard deviation increasing to $\sigma = 0.74$. After sufficient training (Fig.~\ref{fig:fig_8}(c), $\lambda = 1\mathrm{e}{-}3$, epoch = 300), the bimodal pattern becomes more pronounced, the weight density near zero is substantially reduced, and the standard deviation further increases to $\sigma = 0.82$. When the regularization strength further increases (Fig.~\ref{fig:fig_8}(d), $\lambda = 1\mathrm{e}{-}2$, epoch = 300), the weight distribution approaches a near-binary form, with the standard deviation rising to $\sigma = 0.98$. Notably, throughout this process, both the mean and skewness remain close to zero, and the proportion of negative weights remains approximately 50\%, indicating a symmetric distribution without systematic bias. These results demonstrate that NAT reliably transforms the weight distribution from a zero-centered Gaussian to a $\pm 1$ bimodal shape, thereby enhancing the compatibility between quantized weights and eNVM device characteristics.

\begin{table}[t]
\centering
\caption{\textsc{Impact of the regularization coefficient on accuracy (\%)}}
\label{tab:NAT_regularization}
\vspace{-4pt}
\setlength{\tabcolsep}{5pt}
\renewcommand{\arraystretch}{1.70}
\begin{tabular}{
@{}
>{\centering\arraybackslash}p{1.40cm}
*{5}{>{\centering\arraybackslash}p{1.08cm}}
@{}
}
\toprule
\multirow{3}{*}{\textbf{Model}}
& \multicolumn{5}{c}{\textbf{Regularization coefficient ($\lambda$)}} \\
\cline{2-6}
& \multicolumn{4}{c}{\textbf{NAT}}
& \textbf{Baseline} \\
\cmidrule(r{0.5em}){2-5}\cmidrule(l{0.5em}){6-6}
& $10^{-5}$
& $10^{-4}$
& $10^{-3}$
& $10^{-2}$
& $0$ \\
\hline
VGG6
& 91.56 & 93.85 & \textbf{94.68} & 93.25 & 88.65 \\
\hline
VGG11
& 85.40 & 91.82 & \textbf{92.53} & 91.09 & 78.12 \\
\hline
ResNet18
& 60.33 & 66.28 & \textbf{69.16} & 67.64 & 54.36 \\
\bottomrule
\end{tabular}
\vspace{-5pt}
\end{table}


\subsubsection{Evaluation of the Impact of NAT Algorithm Regularization Strength}
This section evaluates the impact of the NAT regularization coefficient $\lambda$ on classification accuracy, where the baseline corresponds to training without NAT (i.e., $\lambda=0$) under identical training configurations. As shown in Table~\ref{tab:NAT_regularization}, as $\lambda$ increases from $10^{-5}$ to $10^{-3}$, the classification accuracy of all three networks improves progressively. However, when $\lambda$ is further increased to $10^{-2}$, the accuracy begins to degrade. The evolution of weight distributions illustrated in Fig.~\ref{fig:fig_8} explains this trend: when $\lambda$ is small, the regularization strength is insufficient and the weights remain predominantly concentrated near zero, causing a substantial portion of weights to be mapped into the intermediate nonlinear conductance region of eNVM devices, where they are susceptible to mapping and update errors. Moderately increasing $\lambda$ guides more weights toward the stable boundary regions near $\pm1$ while retaining a fraction in the intermediate region, thereby mitigating the impact of device non-idealities while preserving the model's representational capacity. However, an excessively large $\lambda$ forces the weights to concentrate overly in the boundary regions, reducing the degrees of freedom for parameter optimization and impairing training convergence. Overall, $\lambda=10^{-3}$ achieves the highest accuracy across all three networks, striking a favorable balance between device robustness and model performance.

\subsection{Evaluation of NAT Algorithm under Device Non-Idealities}
In this section, we evaluate the effectiveness of the NAT algorithm in mitigating the non-idealities of eNVM devices. Based on the device behavior model established in Section \ref{sec:Modeling}, we characterize the nonlinearity exhibited during the LTP and LTD processes using the parameters $(A_{\mathrm{LTP}}/A_{\mathrm{LTD}})$, where larger absolute values correspond to stronger nonlinear deviations. We selected three previously reported devices for comparison: TaO\textsubscript{x}/TiO\textsubscript{2} OxRRAM~\cite{article}, PCMO ReRAM~\cite{8168326}, and IZO FeFET~\cite{abm8537}, along with the CIPS FeFET device fabricated in this work. Table~\ref{tab:Parameters} summarizes the detailed device parameters. By comparing the behavior of these devices within the same simulation framework, we quantitatively assess the robustness and efficacy of NAT in handling a range of device non-idealities.

\setlength{\tabcolsep}{2pt}
\renewcommand{\arraystretch}{1.8}
\begin{table}[t]
\centering
\caption{\textsc{Parameters of different eNVM devices}}
\vspace{-4pt}
\begin{tabular}{lccccc}
\toprule
\multirow{2}{*}{\makecell{\textbf{eNVM}}} &
\multirow{2}{*}{\makecell{\textbf{Nonlinearity} \\
{\scriptsize   $\boldsymbol{(A_{\mathrm{LTP}}/A_{\mathrm{LTD}})}$}}} &
\multirow{2}{*}{\textbf{Ron}} &
\multirow{2}{*}{\makecell{\textbf{ON/OFF} \\ \textbf{ratio}}} &
\multicolumn{2}{c}{\makecell{\textbf{Write pulse}}}  \\
\cline{5-6}
& & & &
{\scriptsize\textbf{Voltage (V)}} & {\scriptsize\textbf{Width (ms)}}  \\
\hline
TaO\textsubscript{x}/TiO\textsubscript{2} & 0.66/-0.69 & 5 M$\Omega$ & 2 & 3/-3 & 40 \\
\hline
PCMO & 3.58/-6.76 & 200 M$\Omega$ & 10 & 4/-3.5 & 10 \\
\hline
IZO & 0.81/-1.15 & 1 M$\Omega$ & 33.1 & 3.96/-3.96 & 10 \\
\hline
\textbf{CIPS/MoS\textsubscript{2}} & 3.2/-4.3 & 9 M$\Omega$ & 3.67 & 1/-1& 110 \\
\bottomrule
\end{tabular}
\label{tab:Parameters}
\vspace{-6pt}
\end{table}

\subsubsection{Evaluation of NAT under Device Nonlinearity and D2D Variations}
\begin{figure*}[t]
    \centering
    \vspace{-3pt}
    \includegraphics[width=1\linewidth]{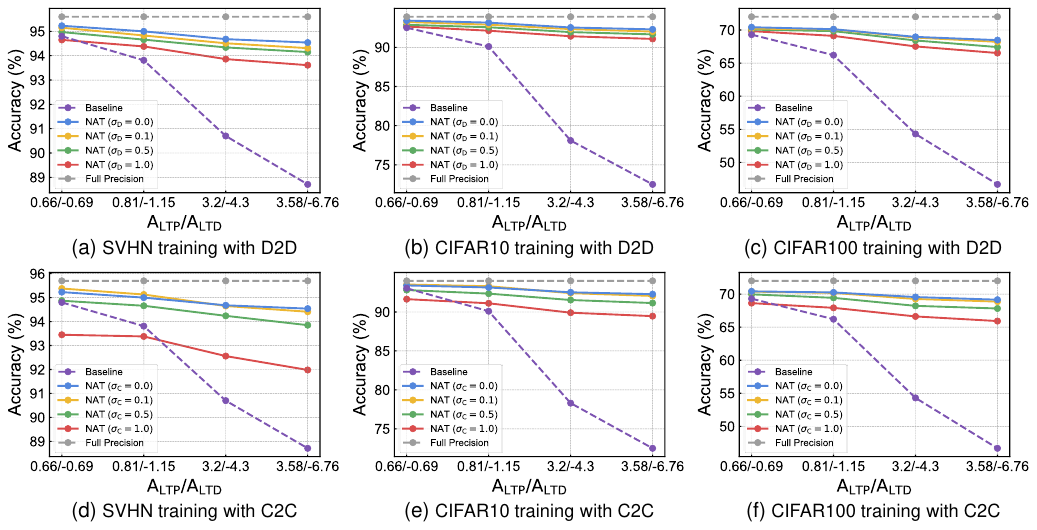}
    \vspace{-18pt}
    \caption{NAT algorithm evaluation results: (a), (b) and (c) performance under varying device nonlinearity and D2D variability; (d), (e) and (f) performance under varying device nonlinearity and C2C variation. Here, the horizontal axis (\(A_{\text{LTP}}/A_{\text{LTD}}\)) represents the nonlinearity of the LTP and LTD phases.}
    \vspace{-4pt}
    \label{fig:fig_9}
\end{figure*}

This section evaluates the performance of the NAT algorithm in mitigating device nonlinearity and D2D variations. Simulations are conducted under varying degrees of device nonlinearity, superimposed with different levels of D2D variations quantified by standard deviations $\sigma_{\text{D}} \in \{0.0, 0.1, 0.5, 1.0\}$. For comparison, full-precision training results are provided as a reference, and training results without NAT under identical settings serve as the baseline. The simulation results are presented in Fig.~\ref{fig:fig_9}(a)--(c).

Experimental results demonstrate that NAT consistently outperforms the conventional training baseline across all evaluated tasks. Specifically, on the VGG6-SVHN task, as the nonlinearity strength increases to (3.58/$-$6.76), the baseline accuracy drops from 94.8\% to 86.71\%, whereas NAT experiences only a marginal decline from 94.68\% to 93.61\%. Similarly, under the strongest nonlinearity condition, NAT maintains accuracies of 91.07\% and 66.52\% on VGG11-CIFAR10 and ResNet18-CIFAR100, respectively, significantly surpassing the baseline models' 72.50\% and 46.68\%. Aggregated across the three tasks under the strongest nonlinearity condition, NAT achieves an average absolute accuracy improvement of 15.1\% over the baseline. Furthermore, even in the absence of D2D variation (i.e., $\sigma_{\mathrm{D}}=0$), the model accuracy still degrades significantly as device nonlinearity increases. This result indicates that the performance degradation of the baseline model primarily stems from device nonlinearity. In contrast, NAT maintains relatively high accuracy even under the most severe nonlinearity conditions, demonstrating its effectiveness in mitigating the impact of device non-idealities.

Further analysis reveals that model accuracy degrades only moderately when D2D variability is introduced. Specifically, under the most severe nonlinearity condition (3.58/-6.76), increasing the D2D noise standard deviation $\sigma_{\text{D}}$ from 0.0 to 1.0 results in only minor performance drops across all tasks: approximately 0.87\% on VGG6 on SVHN, 1.22\% on VGG11 on CIFAR10, and 1.95\% on ResNet18 on CIFAR100. This consistent trend demonstrates that NAT exhibits strong robustness against D2D variations. This behavior stems from the intrinsic mechanism of NAT, which guides weight values during training toward the high and low conductance extremes of the eNVM device range. Physically, eNVM devices exhibit more distinct and stable conductance states at these extreme regions, where the relative magnitude of D2D fluctuations is significantly smaller than in intermediate conductance states. Consequently, mapping weights to these regions effectively reduces the model's sensitivity to fabrication process variations, substantially suppressing the impact of D2D on accuracy. Therefore, NAT not only mitigates mapping errors caused by device nonlinearity but also alleviates the adverse effects of manufacturing variations.

\subsubsection{Evaluation of NAT under Device Nonlinearity and C2C Variations}
This section evaluates the performance of the NAT in handling device nonlinearity and C2C variations. In the simulations, device nonlinearity is characterized by $A_{\mathrm{LTP}}/A_{\mathrm{LTD}}$, upon which C2C noise of varying intensities is superimposed, with standard deviations set to $\sigma_{\mathrm{C}}\in\{0,0.1,0.5,1.0\}$. For comparison, results obtained without NAT under identical training settings serve as the baseline. The corresponding results are presented in Fig.~\ref{fig:fig_9}(d)--(f).

The results indicate that, in the absence of C2C variations ($\sigma_{\mathrm{C}}$ = 0.0), NAT consistently achieves significantly higher classification accuracy than the baseline across all tasks. Notably, under the most severe nonlinearity condition (3.58/-6.76), the NAT-trained models still maintain high accuracy compared to the baseline. These findings validate the effectiveness of NAT in mitigating nonlinear weight updates, enabling the model to achieve accuracy close to that attainable with ideal devices, even when deployed on highly asymmetric non-ideal eNVM hardware.

Meanwhile, the experiments reveal the impact of C2C variations on the performance of the NAT algorithm. As $\sigma_{\mathrm{C}}$ increases, the accuracy of NAT-trained models exhibits only a mild decline. For instance, on VGG11-CIFAR10 under the worst-case nonlinearity (3.58/-6.76), the accuracy is 92.29\% when $\sigma_{\mathrm{C}}$ = 0.0, and decreases to 89.47\% when $\sigma_{\mathrm{C}}$ is increased to 0.5, a drop of only approximately 2.8\%. This phenomenon indicates that, despite the stochastic perturbations introduced by C2C variations during each weight update, which interfere with model convergence, NAT still demonstrates a certain level of robustness against C2C noise.

The underlying mechanism is that NAT pushes the weight distribution toward the extreme HRS and LRS regions of eNVM devices during training. In these regions, the C2C fluctuations have clearer physical boundaries and smaller relative variation, which enables NAT to partially mitigate the destabilizing effect of C2C noise on training stability. Nevertheless, as $\sigma_{\text{C}}$ continues to increase, the accuracy shows a clear downward trend, indicating that C2C noise remains a critical factor limiting the training accuracy. In summary, achieving high accuracy on-chip training requires not only algorithmic compensation mechanisms like NAT but also reliable eNVM with low nonlinearity and minimal C2C variability.

\begin{figure*}[ht]
    \centering
    \includegraphics[width=1\linewidth]{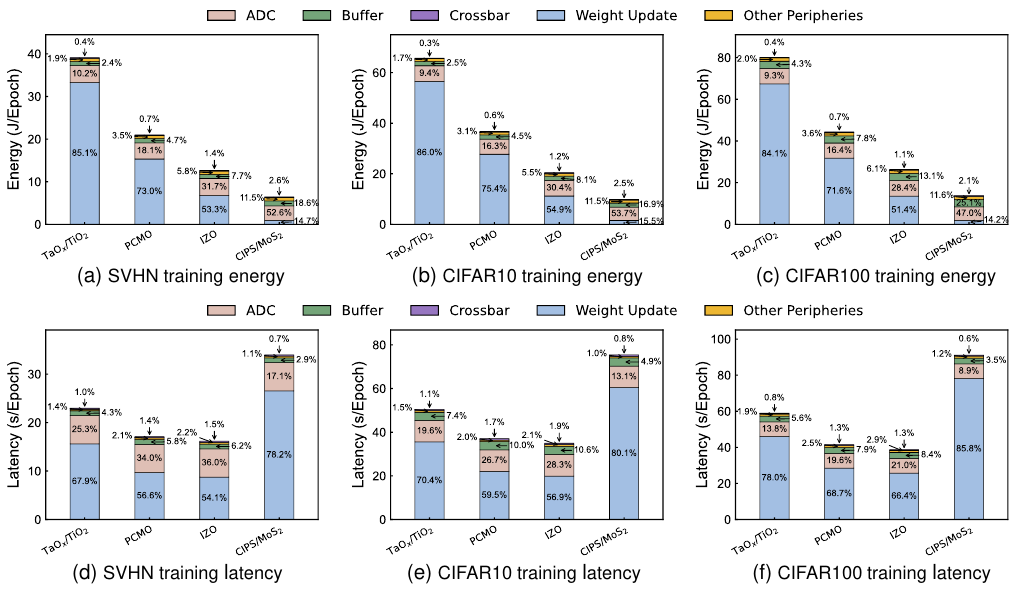}
    \vspace{-18pt}
    \caption{Energy and latency breakdowns of the NOVA architecture across different eNVM device types.}
    \vspace{-9pt}
    \label{fig:fig_10}
\end{figure*}

\subsection{Hardware Performance Evaluation}
\subsubsection{Energy Consumption and Latency Breakdown}
To evaluate the intrinsic relationship between device parameters and system-level performance, this section compares the on-chip training performance of four eNVM devices: TaO\textsubscript{x}/TiO\textsubscript{2}, PCMO, IZO, and CIPS/MoS\textsubscript{2}. Detailed device parameters are listed in Table~\ref{tab:Parameters}. As shown in Fig.~\ref{fig:fig_10}, the energy and latency breakdowns across three experimental configurations indicate that the total energy consumption and latency during the training phase are dominated by weight write operations, while peripheral circuits account for only a negligible fraction. This trend is closely correlated with the write pulse width and voltage characteristics of the different device types.

For resistive-switching devices such as ReRAM, the energy consumption of a single programming pulse is highly dependent on the device conductance, programming voltage, and pulse width. Higher programming voltages or longer pulse widths increase the write energy, introducing greater system-level overhead. In contrast, FeFETs are capacitive devices whose write operation is achieved via ferroelectric polarization switching within the gate dielectric~\cite{abm8537}. Their dominant energy consumption is dictated by the charging and discharging of the equivalent gate capacitance. Therefore, under a simplified capacitive energy model, the write energy of an FeFET is primarily governed by the equivalent capacitance and the programming voltage. Despite their distinct energy mechanisms, the weight update latency for both types of devices is jointly determined by the required number of programming pulses and the duration of a single pulse.

Specifically, TaO\textsubscript{x}/TiO\textsubscript{2} employs a high programming voltage of $\pm 3$~V and a long pulse width of 40~ms, resulting in high energy and latency per update. PCMO reduces the pulse width to 10~ms, lowering energy consumption, but still requires a large number of pulses, limiting the improvement of latency. IZO achieves the shortest training latency across the three tasks due to its superior linearity and tunability, which reduces the required number of pulses even at a 10~ms pulse width. Meanwhile, the CIPS/MoS$_2$ operates at a low voltage of $\pm 1$~V, yielding the lowest write energy and offering significant advantages in energy efficiency. However, its long write pulse width of 110~ms increases the total training time, illustrating a classic trade-off between energy efficiency and speed.

In summary, the analysis of the energy breakdown in Fig.~\ref{fig:fig_10} confirms that weight updates constitute the primary energy bottleneck in eNVM-based accelerators for on-chip training, overwhelmingly dominating the total energy budget. This underscores the critical importance of optimizing the write mechanism—such as reducing programming voltage and shortening pulse width. Notably, the energy consumed by analog computation in the crossbar array is extremely low, fully demonstrating the substantial energy efficiency potential of such architectures as neural network accelerators.

\setlength{\tabcolsep}{5.2pt}
\renewcommand{\arraystretch}{1.9}
\begin{table*}[htbp]
\centering
\caption{\textsc{Performance comparison of different eNVM-based accelerators}}
\vspace{-4pt}
\label{tab:comparison}
\begin{tabular}{cccccccccc}
\toprule
\textbf{\makecell{IMC works}} &
\makecell{\textbf{ISSCC}\textbf{2020}\\~\cite{9062979}} &
\makecell{\textbf{ISSCC}\textbf{2022}\\~\cite{9731725}} &
\makecell{\textbf{VLSI}\textbf{2022}\\~\cite{9830153}} &
\makecell{\textbf{ISSCC}\textbf{2023}\\~\cite{10067544}} &
\makecell{\textbf{N}\textbf{2022}\\~\cite{wan2022compute}} &
\makecell{\textbf{NE}\textbf{2020}\\~\cite{yao2020fully}} &
\makecell{\textbf{ICCAD}\textbf{2024}\\~\cite{3676792}} &
\makecell{\textbf{TCASI}\textbf{2025}\\~\cite{10753271}} &
\makecell{\textbf{NOVA}\\[2pt]\textbf{(This work)}} \\
\hline
\textbf{Cell structure} & 1T1R & 1T1R & 1T1M & 1T1R+SRAM & 1T1R & 1T1R & 1T2F & 1T1R & 1T1F \\
\hline
\textbf{Memory type} & ReRAM & ReRAM & MRAM & ReRAM/SRAM & ReRAM & ReRAM & FeFET & ReRAM & FeFET \\
\hline
\textbf{Technology} & 130nm & 40nm & 22nm & 40nm & 130nm & 130nm & 45nm & 28nm & 32nm \\
\hline
\textbf{ADC precision} & N/A & 4b & 6b & 6b & N/A & 8b & 5b & 8b & 6b \\
\hline
\textbf{Area (mm$^2$)} & 1.79 & 0.027 & 0.225 & 20.25 & 159 & 0.0704 & 232.9 & N/A & 43.75 \\
\hline
\textbf{Power (mW)} & 2.2 & 0.113 & N/A & 21.3 & N/A & 7.438 & N/A & \makecell{72.4 (FP)\\80.4 (BP)} & \makecell{18.23 (FP)\\20.59 (BP)} \\
\hline
\textbf{Leakage power} & 17uW & N/A & N/A & 0.43mW & N/A & N/A & 1.54mW & N/A & 0.368mW \\
\hline
\textbf{\makecell{Precision\\(input, weight)}} & (1b, 4b) & (1b, 1b) & (1b, 4b) & (1b, 1b) & (4b, 4b) & (8b, 8b) & (8b, 8b) & (8b, 8b) & (8b, 8b) \\
\hline
\textbf{\makecell{Energy efficiency\\(TOPS/W)}} & 74 & 26.56 & 41.6 & 73.53 & 16 & 11.01 & 48.03 & \makecell{3.58 (FP)\\3.26 (BP)} & \makecell{51.08 (FP)\\47.63 (BP)} \\
\hline
\textbf{Application} & \textbf{Training} & Inference & Inference & Inference & \textbf{Training} & Inference & Inference & \textbf{Training} & \textbf{Training} \\
\bottomrule
\end{tabular}
\end{table*}

\begin{figure}[ht]
    \centering
    \includegraphics[width=0.8\linewidth]{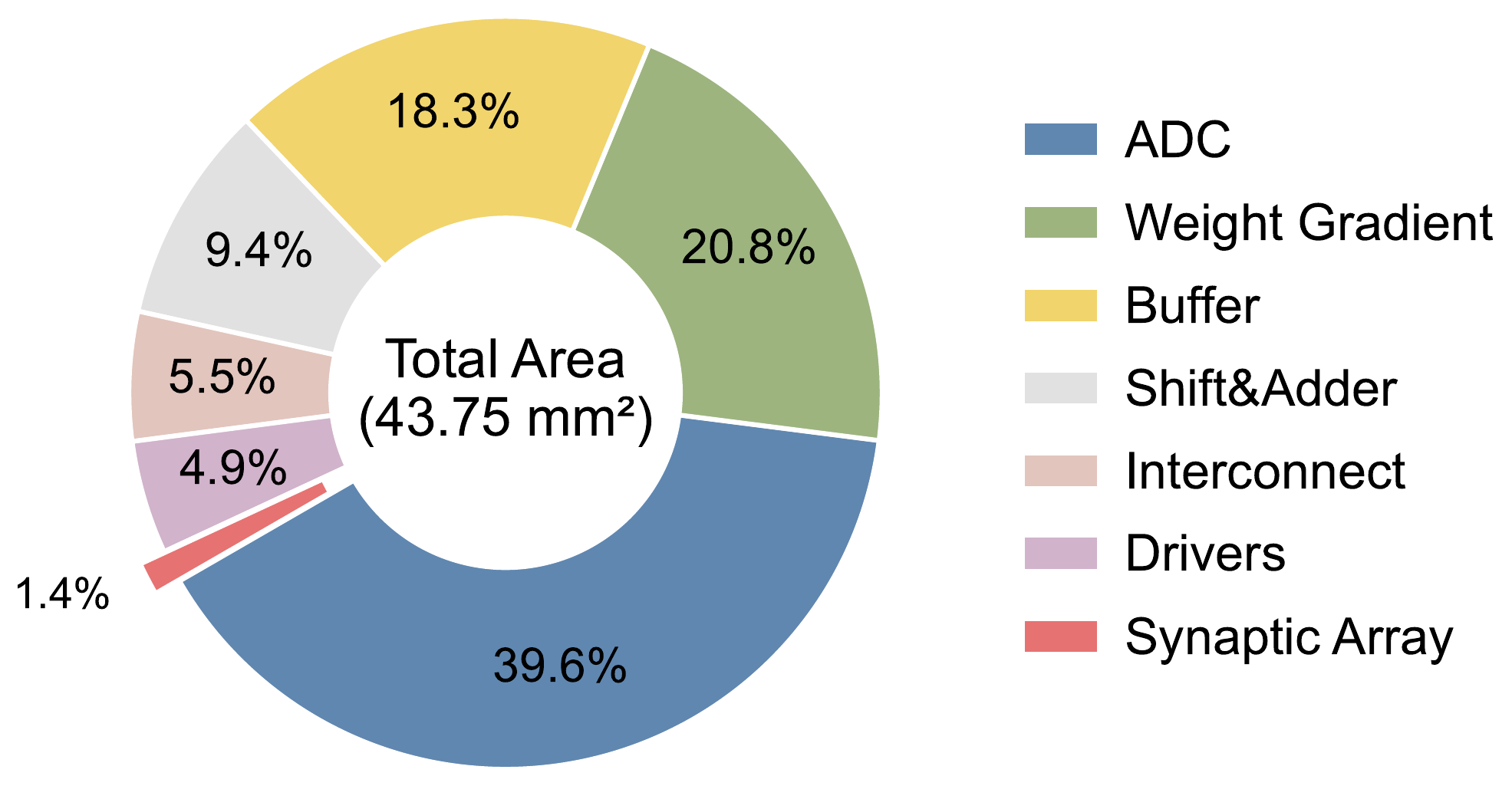}
    \caption{Area breakdown of the NOVA architecture.}
    \label{fig:fig_11} 
\end{figure}

\subsubsection{Area Breakdown}
Fig.~\ref{fig:fig_11} presents the area breakdown of the NOVA architecture, with a total system area of 43.75$~\mathrm{mm}^2$. The ADCs account for the largest proportion, at 39.6\%, indicating that the 6-bit ADCs remain the primary source of system area overhead. The weight-gradient computation unit accounts for 20.8\% of the total area, demonstrating that explicit weight-gradient computation introduces non-negligible hardware overhead. Buffers account for 18.3\%, reflecting the substantial on-chip storage resources required to retain intermediate activations, errors, gradients, and partial sums during on-chip training. The Shift\&Adder units, on-chip interconnects, and drivers account for 9.4\%, 5.5\%, and 4.9\% of the total area, respectively. In contrast, the FeFET synaptic arrays, which serve as the core compute and storage units, occupy only 1.4\% of the total area, highlighting the high area efficiency of crossbar arrays. Notably, although the ADCs and buffers occupy a large fraction of the chip area, their contributions to the training energy are relatively low because they are activated only during readout. By contrast, on-chip training requires frequent high-voltage, long-pulse programming operations to adjust the conductance states of the eNVM devices, making conductance updates the dominant source of system energy consumption.

\subsection{Related Work and Comparison}
\subsubsection{Comparison with Existing eNVM Non-Ideality Mitigation Techniques}
Existing approaches to mitigating eNVM device non-idealities can be broadly categorized into off-chip hardware-aware training and on-chip training. Off-chip methods improve post-deployment inference accuracy by injecting device non-idealities during training~\cite{rasch2023hardware}, performing adaptive quantization based on per-cell conductance sensing~\cite{adp3710}, or employing programming-aware retraining~\cite{10323793}. However, backpropagation and parameter updates in these methods are still performed by off-chip software. For on-chip training, the work in~\cite{li2018efficient} employs a two-pulse conductance tuning scheme to improve weight update linearity, but its backpropagation still relies on external software, preventing complete end-to-end on-chip training. The work in~\cite{yi2023activity} computes weight updates by comparing the local activity differences between the free and clamped phases of the network, eliminating the need for explicit backpropagation. However, this method requires converting the feedforward network into a bidirectional Hopfield-type network and redundantly mapping both the weight matrix and its transpose in a sparse manner, which inevitably increases storage and mapping overhead. The work in~\cite{9292971} employs momentum-based updates to mitigate write errors, but its effectiveness remains constrained by device update linearity, leading to convergence difficulties and accuracy degradation under severe nonlinearity.

Overall, existing on-chip training approaches remain limited by incomplete training loops and a strong dependence on device-update linearity. In contrast, NOVA differs fundamentally in both its training execution mechanism and non-ideality mitigation strategy. At the hardware level, NOVA employs peripheral circuitry with configurable dataflow to perform forward propagation, backpropagation, and weight updates entirely within the eNVM accelerator, achieving complete end-to-end on-chip training. At the algorithmic level, NOVA jointly optimizes the proposed NAT regularization term and task loss to actively reduce the weight density in highly nonlinear intermediate-conductance regions and guide weights toward relatively stable and approximately linear boundary regions, ultimately minimizing weight mapping errors. Experimental results demonstrate that, compared with the existing method in~\cite{9292971}, NOVA achieves an average accuracy improvement of 15.1\% across multiple networks and datasets.

\subsubsection{Hardware Performance Comparison}
As shown in Table~\ref{tab:comparison}, although NOVA targets more complex on-chip training tasks, it still achieves a forward-propagation energy efficiency of 51.08 TOPS/W at (8b, 8b) input and weight precision. Under the same precision configuration, this metric exceeds the 11.01 TOPS/W and 48.03 TOPS/W reported by the inference-oriented IMC accelerators in~\cite{yao2020fully} and~\cite{3676792}, respectively. Compared with designs that primarily employ low-bit-width configurations, including those in~\cite{9731725,9830153,10067544}, NOVA maintains competitive forward-propagation energy efficiency while supporting higher numerical precision and a complete training dataflow. Furthermore, compared with eNVM accelerators designed for on-chip training, NOVA achieves forward- and backward-propagation energy efficiencies of 51.08 TOPS/W and 47.63 TOPS/W, respectively, at (8b, 8b) precision. Relative to low-bit-width on-chip training designs~\cite{9062979,wan2022compute}, NOVA remains highly competitive in energy efficiency while supporting higher input and weight precision. In particular, its forward- and backward-propagation energy efficiencies are 14.27$\times$ and 14.61$\times$ those of the design in~\cite{10753271}, respectively, at the same (8b, 8b) precision, demonstrating a favorable balance between numerical precision and training energy efficiency.

To further quantify the energy-efficiency advantage of NOVA over GPU platforms, we adopt the peak INT8 Tensor Core energy efficiency of the NVIDIA RTX 4090 GPU as the reference baseline. According to NVIDIA's official technical documentation~\cite{nvidia_ada_whitepaper}, the theoretical peak energy efficiency of the RTX 4090 at INT8 precision is approximately 1.47 TOPS/W. In contrast, NOVA achieves forward- and backward-propagation energy efficiencies of 51.08 TOPS/W and 47.63 TOPS/W, respectively, during on-chip training, corresponding to an average energy efficiency of 49.36 TOPS/W. Even when the theoretical peak energy efficiency of the GPU is used as a stringent reference, NOVA achieves an average energy efficiency approximately 33.58$\times$ that of the GPU, further demonstrating the significant energy-efficiency potential of eNVM-based IMC architectures for on-chip training.

\section{Conclusion}
\label{sec:section_5}
In this work, we present an on-chip training architecture based on eNVM. First, we fabricated CIPS/MoS$_2$ heterojunction-based FeFET devices and performed parameter fitting and behavioral modeling of their electrical characteristics using measured data.
Building on this characterization, we further developed a training algorithm designed to mitigate accuracy degradation caused by eNVM device non-idealities. Experimental results demonstrate that even under highly asymmetric device characteristics, the proposed NAT algorithm achieves training accuracy comparable to that obtained with ideal devices. Furthermore, compared with existing advanced methods, the proposed design exhibits significant energy efficiency advantages. Collectively, these results indicate that the proposed architecture provides a practical solution for achieving high-performance on-chip training with eNVM technology.

\raggedbottom
\printbibliography[title={R\textsc{eferences}}]
\end{document}